\documentclass[conference]{IEEEtran}
\IEEEoverridecommandlockouts
\usepackage{cite}
\usepackage[ruled,linesnumbered]{algorithm2e}
\usepackage{amsmath,amssymb,amsfonts}
\usepackage{algorithmic}
\usepackage{graphicx}
\usepackage{textcomp}
\usepackage{graphicx}
\usepackage{subcaption}  
\usepackage{xcolor}
\usepackage{float}
\usepackage{dblfloatfix}
\usepackage[hang,flushmargin]{footmisc}

\def\BibTeX{{\rm B\kern-.05em{\sc i\kern-.025em b}\kern-.08em
		T\kern-.1667em\lower.7ex\hbox{E}\kern-.125emX}}
\newcommand{\mv}[1]{\mbox{\boldmath{$ #1 $}}}
\usepackage[T1]{fontenc}
\usepackage{times}
\usepackage[T1]{fontenc}
\begin{document}
\title{Sequential Transmit Covariance Optimization for Wireless Tracking Exploiting Observation History}

\author{\IEEEauthorblockN{Kaiyue Hou and Shuowen Zhang}
\IEEEauthorblockA{{Department of Electrical and Electronic Engineering, The Hong Kong Polytechnic University} \\
E-mail: kaiyue.hou@connect.polyu.hk, shuowen.zhang@polyu.edu.hk}
}
\maketitle
\begin{abstract}
This paper studies a multiple-input multiple-output (MIMO) radar tracking system, where a multi-antenna base station (BS) aims to track the location of a moving target over multiple time slots based on the observed echo signals, initial prior probability density function (PDF) of the target state, and state evolution model. By exploiting the realized observation history in the past time slots, the BS sequentially updates the predictive state information and designs the transmit covariance matrix before collecting the current echo observation. Considering a Gaussian random-walk model for the target location and a Gauss-Markov model for the complex radar cross-section (RCS) coefficient, we propose an effective method to characterize the predictive PDF conditioned on the realized observation history via Gaussian approximation. Based on this, we derive the conditional posterior Fisher information matrix (PFIM) for the target state, and further characterize the conditional posterior Cramér-Rao bound (PCRB) for the mean-squared error (MSE) in estimating the target's location state as an explicit expression of the transmit covariance matrix. Next, we formulate the sequential transmit covariance matrix optimization problem to minimize the conditional PCRB for each time slot. Despite the non-convexity of the problem, we obtain the optimal solution via the Schur complement technique. Numerical results show that the proposed design effectively exploits the realized observation history, achieves a lower conditional PCRB than the benchmark schemes, and improves tracking accuracy over time.  
\end{abstract}

\section{Introduction}
Transmit signal design is of significant importance to radar and integrated sensing and communication (ISAC) systems as it determines the quality of received echo signals (i.e., observations) used for sensing. Over multiple time slots, the transmitter can further sequentially adapt its probing strategy to the observation history to enhance sensing performance. This is particularly relevant to wireless tracking, since the target's state (e.g., location) also evolves over time, and the sensing performance is jointly determined by the state evolution, observation uncertainty, and transmit signal designs.

Posterior Cram\'er-Rao bound (PCRB), also called Bayesian Cram\'er-Rao bound (BCRB), is a global lower bound of the estimation mean-squared error (MSE) and is independent of the exact values of the parameters to be estimated \cite{Tichavsky1998PCRB}, \cite{bibTrees}, thereby being a viable sensing performance metric for transmit signal design before sensing is performed. For \emph{static} parameter estimation, transmit signals can be designed based on the prior probability density function (PDF) of the static parameter to minimize the PCRB, such that the posterior PDF is shaped to favor sensing \cite{xu2024mimo,Hou_JSAC,yao2025beamforming,ZhuTWC_2025_Fronthaul}. On the other hand, for \emph{non-static time-varying} parameter estimation such as in the tracking of moving targets, transmit signals in each time slot need to be designed based on a ``predictive PDF'' dependent on both the observation history and the state evolution model over time, thus being even more challenging. Particularly, for location tracking, the PCRB conditioned on the observation history is very difficult to characterize due to the complex function form of the observations (echo signals) with respect to the state (location). Hence, initial studies generally characterized the conditional PCRB \emph{numerically} via Monte Carlo methods, particle filtering, and/or recursive methods \cite{hurtado2008adaptive,sharaga2015optimal}, which are computationally extensive especially with long observation history and cannot reveal the explicit relationship between the PCRB and transmit signals. There has also been another line of works on tracking that employed data-driven learning approaches (see, e.g., \cite{han2024active}). To the best of the authors' knowledge, \emph{analytical} and \emph{explicit} characterization of the wireless radar tracking performance exploiting observation history together with the \emph{sequential transmit signal optimization} based upon it remain open problems for investigation.

\begin{figure}[t]
\centering
\includegraphics[width=0.89\linewidth]{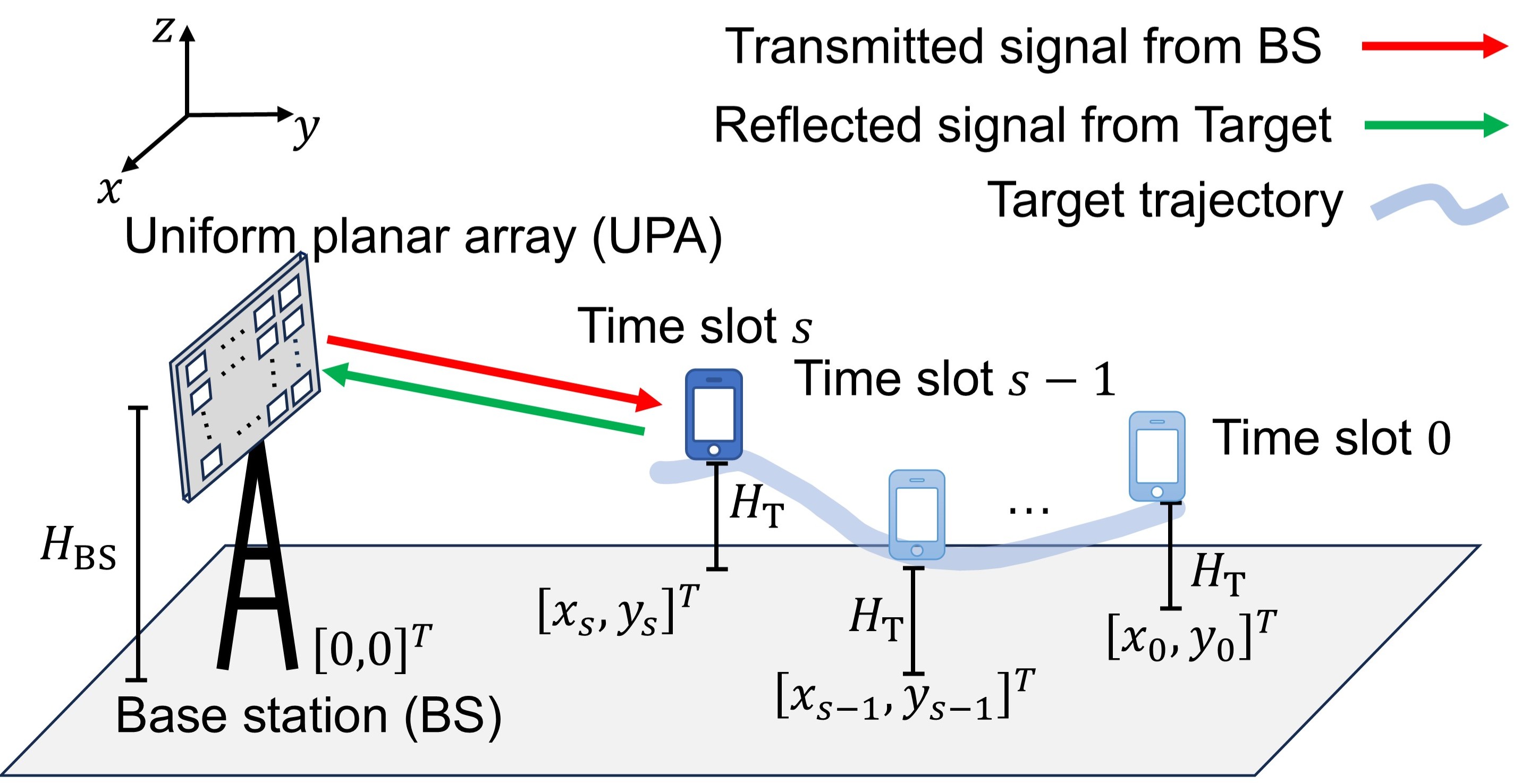}
\setlength{\abovecaptionskip}{-0.1cm}\vspace{2mm}
\caption{Illustration of a MIMO radar tracking system.}
\label{Fig1_system}\vspace{-2mm}
\end{figure}

Motivated by the above, this paper studies explicit tracking performance characterization and sequential transmit signal optimization for a MIMO radar tracking system, where a multi-antenna base station (BS) aims to track the location of a moving point target over multiple time slots from the observed echo signals, initial prior PDF of the target's location, and the state evolution model. In each time slot, the BS constructs a predictive PDF and designs the transmit signals before collecting the new echo observations. With the time-varying target location modeled as a Gaussian random-walk process and the complex radar cross-section (RCS) coefficient modeled as a Gauss-Markov process, we construct an assumed-density conditional posterior Fisher information matrix (PFIM) for their joint state. Treating the RCS coefficient as a nuisance parameter, we derive the conditional PCRB for the target location as an explicit function of the transmit covariance matrix. Then, we obtain the optimal transmit covariance matrix in each time slot that minimizes the conditional PCRB, despite the non-convexity of the original problem. It is shown via numerical results that the proposed design achieves improving tracking performance over time and outperforms various benchmark schemes due to the effective exploitation of the observation history.

\section{System Model}\label{sec:system_model}
Consider a MIMO radar tracking system as illustrated in Fig. \ref{Fig1_system}, where a multi-antenna BS aims to track the horizontal location of a moving point target with a known fixed height $H_{\rm T}$ in a three-dimensional (3D) space. The BS is located at a known location denoted by $\mv b\triangleq[0,0,H_{\rm BS}]^T$, where $H_{\rm BS}$ denotes the height of the BS. The BS is equipped with $N_{\rm T}=N_{{\rm T},x}N_{{\rm T},z}$ transmit antennas and $N_{\rm R}=N_{{\rm R},x}N_{{\rm R},z}$ receive antennas both configured as uniform planar arrays (UPAs). Both UPAs are placed on the $x$-$z$ plane with broadside directions along the $y$-axis. Tracking is performed over $S\geq 1$ discrete time slots, each with time duration $T_{\rm s}$ consisting of $L$ sample intervals. Let $s\in\{0,1,\ldots,S\}$ denote the time slot index, where the $s$-th time slot corresponds to time interval $[sT_{\rm s},(s+1)T_{\rm s}]$. The target location and RCS coefficient are assumed to remain constant within each time slot, and evolve over different time slots according to evolution models below.

\subsection{Target Location and RCS Coefficient Evolution Models}
The time-varying horizontal location of the target is modeled as a discrete-time state sequence $\{\mv u_s\}_{s=0}^{S}$, where $\mv u_s\triangleq[x_s,y_s]^T$ denotes the unknown horizontal coordinates of the target in the $s$-th time slot, as illustrated in Fig.~\ref{Fig1_system}. Since the target has a known fixed height $H_{\rm T}$, only its horizontal coordinates are \textit{unknown} and \textit{time-varying}. The 3D target location in time slot $s$ is given by $[x_s,y_s,H_{\rm T}]^T$. The initial horizontal location of the target, $\mv u_0\in\mathbb R^2$, is modeled as a Gaussian random vector with prior PDF $p_{\mv U_0}(\mv u_0)=\frac{1}{2\pi\det(\mv C_0)^{1/2}}e^{-\frac{1}{2}(\mv u_0-\bar{\mv u}_0)^T\mv C_0^{-1}(\mv u_0-\bar{\mv u}_0)}$, 
where $\bar{\mv u}_0$ denotes the mean of the initial target horizontal location, and $\mv C_0\succ\mv 0$ denotes the covariance matrix that characterizes the initial location uncertainty. The target motion is assumed to follow a discrete-time Gaussian random-walk model \cite{gustafsson2005mobile} given by
\begin{align}
\mv u_s=\mv u_{s-1}+\mv\delta_{s-1},\quad s=1,\ldots,S,
\label{motion_model}
\end{align}
where $\mv\delta_{s-1}\sim\mathcal{N}(\mv 0,\mv Q_{s-1})$ denotes the process noise that captures the random horizontal displacement during the time interval $[(s-1)T_{\rm s},sT_{\rm s}]$, and $\mv Q_{s-1}\succeq\mv 0$ denotes the corresponding motion covariance matrix. 

The complex RCS coefficient $\alpha_s$ captures the amplitude and phase change of the signal via target reflection, which depends on the properties of the target's surface facing the incident wireless signals. The initial RCS coefficient is assumed to follow a zero-mean complex Gaussian model given by $\alpha_0\sim\mathcal{CN}(0,\sigma_\alpha^2)$, which holds for Swerling-I and Swerling-II models. $\alpha_s$ is assumed to evolve according to a first-order Gauss--Markov process given by 
\begin{align}
\alpha_s=\rho\alpha_{s-1}+n_{\alpha,s-1},\quad s=1,\ldots,S,
\label{eq:alpha_GM}
\end{align}
where $n_{\alpha,s-1}\sim\mathcal{CN}(0,(1-\rho^2)\sigma_\alpha^2)$ denotes the innovation (residual) noise, and $\rho\in[0,1)$ denotes the temporal correlation coefficient. The innovation noise sequence $\{n_{\alpha,s}\}_{s=0}^{S-1}$ is assumed to be independent across time and independent of the target motion and receiver noise. Under this setting, $\{\alpha_s\}_{s=0}^{S}$ is a zero-mean Gauss--Markov process with marginal distribution $\alpha_s\sim\mathcal{CN}(0,\sigma_\alpha^2)$ for all $s$. Therefore, its average power $\mathbb E[|\alpha_s|^2]=\sigma_\alpha^2$ remains constant over time, while its temporal correlation is determined by $\rho$.

\subsection{Signal Model for Tracking}
The objective of the MIMO radar tracking system is to estimate and track the \emph{time-varying horizontal location} of the target, $\{\mv{u}_s\}_{s=0}^S$, based on the target-reflected echo signals received back at the BS as well as the initial prior distributions and the evolution models introduced above.

In each time slot $s$, the BS-target distance is given by $d_s\triangleq\left\|[x_s,y_s,H_{\rm T}]^T-\mv b\right\|$. The azimuth and elevation angles of the target are given by $\arccos\chi_s\triangleq\arccos\frac{x_s}{d_s}$ and $\arccos\zeta_s\triangleq\arccos\frac{H_{\rm T}-H_{\rm BS}}{d_s}$, respectively. Assuming half-wavelength antenna spacing, the UPA array response of the BS transmitter and BS receiver is given by $\mv a_q(\chi_s,\zeta_s)=\mv a_{q,z}(\zeta_s)\otimes\mv a_{q,x}(\chi_s)$, where
$\mv a_{q,x}(\chi_s)\triangleq\frac{1}{\sqrt{N_{q,x}}}\big[1,e^{-j\pi\chi_s},\ldots,e^{-j\pi(N_{q,x}-1)\chi_s}\big]^T$, 
$\mv a_{q,z}(\zeta_s)\triangleq\frac{1}{\sqrt{N_{q,z}}}\big[1,e^{-j\pi\zeta_s},\ldots,e^{-j\pi(N_{q,z}-1)\zeta_s}\big]^T$, $q\in\{{\rm R},{\rm T}\}$ with $q={\rm T}$ and $q={\rm R}$ corresponding to the transmit and receive arrays, respectively. Let $\beta_0$ denote the reference channel power gain at distance $1$~m. The overall complex round-trip channel gain in time slot $s$ is then modeled as $\beta_s\triangleq\frac{\beta_0}{d_s^2}\alpha_s$. 
The overall channel from the BS transmitter to the BS receiver via target reflection in time slot $s$ is given by
\begin{align}
\mv H_s=\beta_s\mv a_{\rm R}(\chi_s,\zeta_s)\mv a_{\rm T}^H(\chi_s,\zeta_s).
\end{align}
Note that $d_s$, $\chi_s$, $\zeta_s$, $\alpha_s$, $\beta_s$, and consequently $\mv H_s$ remain constant within each time slot $s$ containing $L$ sample intervals.

Let $\mv s_{s,\ell}\in\mathbb C^{N_{\rm T}\times 1}$ denote the baseband equivalent signal vector transmitted in the $\ell$-th sample interval of time slot $s$, where $\ell=1,\ldots,L$. The transmit sample covariance matrix in time slot $s$ is defined as
\begin{align}
\mv W_s\triangleq\frac{1}{L}\sum_{\ell=1}^{L}\mv s_{s,\ell}\mv s_{s,\ell}^{H},\quad s=0,1,\dots,S.
\label{eq:transmit_sample_covariance}
\end{align}
Let $P$ denote the transmit power constraint, which yields ${\rm tr}(\mv W_s)\leq P$. The received baseband echo signal at the BS in the $\ell$-th symbol interval of time slot $s$ is given by
\begin{align}
\hspace{-2mm}\mv y_{s,\ell}\!=\!\mv H_s\mv s_{s,\ell}\!+\!\mv n_{s,\ell}\!=\!\beta_s\mv a_{\rm R}(\chi_s,\zeta_s)\mv a_{\rm T}^H(\chi_s,\zeta_s)\mv s_{s,\ell}\!+\!\mv n_{s,\ell},\!\!
\end{align}
where $\mv n_{s,\ell}\sim\mathcal{CN}(\mv 0,\sigma^2\mv I_{N_{\rm R}})$ denotes the circularly symmetric complex Gaussian (CSCG) receiver noise independent across sample intervals. 

For notational convenience, we define the stacked received signal vector in time slot $s$ as
\begin{align}
\mv y_s
\triangleq
\big[\mv y_{s,1}^T,\mv y_{s,2}^T,\ldots,\mv y_{s,L}^T\big]^T
\in\mathbb C^{LN_{\rm R}\times 1},
\end{align}
and the stacked observation vector up to time slot $s$ as
\begin{align}
\mv y_{0:s}
\triangleq
\big[\mv y_0^T,\mv y_1^T,\ldots,\mv y_s^T\big]^T
\in\mathbb C^{(s+1)LN_{\rm R}\times 1}.
\end{align}

\subsection{Overall Tracking Procedure}     
The overall tracking procedure operates as follows. At the start of each time slot $s$, the BS designs its transmit signals based on the \emph{previous observation history} $\mv y_{0:s-1}$ collected in the previous time slots (if any) as well as the initial prior distributions and the evolution models. The BS then sends the designed signals over the $L$ sample intervals and collects new echo observation $\mv y_s$ in the current time slot. Based on the updated observation history $\mv y_{0:s}=[\mv y_{0:s-1}^T,\mv y_s^T]^T$, the BS estimates the current target location $\mv u_s$. The updated observation history is then used for the signal design in the $(s+1)$-th time slot, so on and so forth. In the following, we characterize the tracking performance, based on which we will study the transmit signal optimization.
\vspace{-1mm}
\section{Tracking Performance Characterization via Conditional PCRB}\label{PCRB}
\vspace{-1mm}
We characterize the tracking performance in each time slot via the conditional PCRB for the target horizontal location estimation MSE, which is conditioned on the previous observation history. 
Note that both the target horizontal location $\mv{u}_s$ and the complex RCS coefficient $\alpha_s$ are unknown parameters. Define $\boldsymbol\alpha_s=[\alpha_s^{\rm R},\alpha_s^{\rm I}]^T$ where $\alpha_s^{\rm R}\triangleq\Re\{\alpha_s\}$ and $\alpha_s^{\rm I}\triangleq\Im\{\alpha_s\}$ denote the real and imaginary parts of $  \alpha_s$, respectively. The overall tracking state at each $s$-th time slot can be defined as
\begin{align}
\mv w_s\triangleq\big[\mv u_s^T,\boldsymbol\alpha_s^T\big]^T\in\mathbb R^{4\times1}.
\label{eq:def-wt}
\end{align}

Conditioned on the observation history $\mv y_{0:s-1}$, let $\mv J_{s\mid 0: s-1}\in\mathbb R^{4\times4}$ denote the conditional PFIM for estimating the target state $\mv w_s$, which is defined as
\begin{align}
\hspace{-3mm}\mv J_{s\mid 0:  s-1}\triangleq
-\mathbb E_{\mv w_s,\mv y_s\mid\mv y_{0:s-1}}\!\!\big[\nabla_{\mv w_s}^{2}\log p(\mv w_s,\mv y_s\mid\mv y_{0:s-1})\big].\!\!\!
\label{eq:conditional_PFIM_definition}
\end{align}
Based on $p(\mv w_s,\mv y_s\mid\mv y_{0:s-1})=p(\mv y_s\mid\mv w_s,\mv y_{0:s-1})p(\mv w_s\mid\mv y_{0:s-1})$, 
the conditional PFIM can be decomposed as
\begin{align}
\mv J_{s\mid0:s-1}=\mv J_{s\mid0:s-1}^{(\rm D)}+\mv J_{s\mid0:s-1}^{(\rm P)},
\label{eq:J_overall_decomp}
\end{align}
where
\begin{align}
\hspace{-3mm}\mv J_{s\mid0:s-1}^{(\rm D)}\!
\triangleq\!-\mathbb E_{\mv w_s,\mv y_s\mid\mv y_{0:s-1}}\!\big[\nabla_{\mv w_s}^{2}\log p(\mv y_s\mid\mv w_s,\mv y_{0:s-1})
\big],\!\!\!\label{eq:J_data_def}
\end{align}
denotes the data information from the current echo observation $\mv y_s$, and
\begin{align}
\hspace{-3mm}\mv J_{s\mid0:s-1}^{(\rm P)}
\triangleq-\mathbb E_{\mv w_s\mid \mv y_{0:s-1}}\!\big[\nabla_{\mv w_s}^{2}\log p\big(\mv w_s\mid \mv y_{0:s-1}\big)
\big],
\label{eq:J_prior_def}
\end{align}
denotes the prior information contained in the predictive PDF $p(\mv w_s\mid\mv y_{0:s-1})$. For $s\geq1$, based on the target motion model in \eqref{motion_model} and the RCS coefficient evolution model in \eqref{eq:alpha_GM}, the unknown tracking state evolves according to 
\begin{align}
\mv w_s
=
\mv F_{s-1}\mv w_{s-1}
+
\mv q_{s-1}, \quad s\geq 1,
\label{eq:motion_state}
\end{align}
where $\mv F_{s-1}\triangleq{\rm blkdiag}(\mv I_2,\rho\mv I_2)$ denotes the state transition matrix, $\mv q_{s-1}\sim\mathcal N(\mv 0,\mv Q^w_{s-1})$ denotes the process noise, and $\mv Q^w_{s-1}\triangleq {\rm blkdiag}\big(\mv Q_{s-1},\frac{(1-\rho^2)\sigma_\alpha^2}{2}\mv I_2\big)$ denotes its covariance matrix. Accordingly, the state transition PDF is given by
\begin{align}
p(\mv w_s\mid\mv w_{s-1})
=\mathcal N\!\big(\mv w_s;\mv F_{s-1}\mv w_{s-1},\mv Q^w_{s-1}\big).
\label{eq:state_transition_density}
\end{align}
The exact predictive PDF for $s\geq 1$ is given by
\begin{align}
\hspace{-3mm}p(\mv w_s\mid \mv y_{0:s-1})
\!\!=\!\!\!
\int\!\!\!
p(\mv w_s\mid\mv w_{s-1})
p(\mv w_{s-1}\mid \mv y_{0:s-1})
d\mv w_{s-1}.\!\!\!\!
\label{eq:predictive_density_exact}
\end{align}
Due to the nonlinearity in the observation model, the posterior PDF $p(\mv w_{s-1}\mid\mv y_{0:s-1})$ and consequently the predictive PDF in \eqref{eq:predictive_density_exact} are generally unavailable in closed form. This motivates us to first derive the conditional data PFIM for a general predictive PDF and then introduce a Gaussian assumed density approximation for tractable evaluation.     
\vspace{-2mm}
\subsection{Conditional Data PFIM}
\vspace{-2mm}
We derive the data information from the current echo observation $\mv y_s$, conditioned on the observation history $\mv y_{0:s-1}$. Conditioned on $\mv y_{0:s-1}$, the transmit signals $\{\mv s_{s,\ell}\}_{\ell=1}^{L}$ have been determined before $\mv y_s$ is collected and are known at the BS receiver. The conditional likelihood is given by
\begin{align}
p(\mv y_s\mid \mv w_s,\mv y_{0:s-1})=\mathcal{CN}\!\big(\boldsymbol\mu_s(\mv w_s),\sigma^2\mv I_{LN_{\rm R}}\big),
\end{align}
where
$
\boldsymbol{\mu}_s(\mv w_s)
=
\begin{bmatrix}
\boldsymbol{\mu}_{s,1}^T(\mv w_s), &
\cdots, &
\boldsymbol{\mu}_{s,L}^T(\mv w_s)
\end{bmatrix}^T$, with
$\boldsymbol{\mu}_{s,\ell}(\mv w_s)
=\frac{\beta_0\alpha_s}{d_s^2}
\mv a_{\rm R}(\chi_s,\zeta_s)\mv a_{\rm T}^H(\chi_s,\zeta_s)
\mv s_{s,\ell},
\quad \ell=1,\ldots,L$.

Next, we derive the required derivatives of $\boldsymbol{\mu}_{s,\ell}(\mv w_s)$. From the UPA responses defined in Section \ref{sec:system_model}, we have
$\frac{\partial\mv a_q(\chi_s,\zeta_s)}{\partial \chi_s}
=\mv a_{q,z}(\zeta_s)\otimes
\frac{\partial\mv a_{q,x}(\chi_s)}{\partial \chi_s}$ and
$\frac{\partial\mv a_q(\chi_s,\zeta_s)}{\partial \zeta_s}
=\frac{\partial\mv a_{q,z}(\zeta_s)}{\partial \zeta_s}\otimes
\mv a_{q,x}(\chi_s)$, where
$\frac{\partial\mv a_{q,x}(\chi_s)}{\partial \chi_s}
=-j\pi\operatorname{diag}(0,1,\ldots,N_{q,x}-1)\mv a_{q,x}(\chi_s)$ and
$\frac{\partial\mv a_{q,z}(\zeta_s)}{\partial \zeta_s}
=-j\pi\operatorname{diag}(0,1,\ldots,N_{q,z}-1)\mv a_{q,z}(\zeta_s)$,
for $q\in\{{\rm R},{\rm T}\}$. The operator $\otimes$ denotes the Kronecker product. For ease of exposition, we define
$\mv B_s\triangleq
\mv a_{\rm R}(\chi_s,\zeta_s)\mv a_{\rm T}^H(\chi_s,\zeta_s)$,
$\mv B_{s,\chi}\triangleq
\frac{\partial\mv a_{\rm R}(\chi_s,\zeta_s)}{\partial \chi_s}
\mv a_{\rm T}^H(\chi_s,\zeta_s)
+\mv a_{\rm R}(\chi_s,\zeta_s)
\frac{\partial\mv a_{\rm T}^H(\chi_s,\zeta_s)}{\partial \chi_s}$, and
$\mv B_{s,\zeta}\triangleq
\frac{\partial\mv a_{\rm R}(\chi_s,\zeta_s)}{\partial \zeta_s}
\mv a_{\rm T}^H(\chi_s,\zeta_s)
+\mv a_{\rm R}(\chi_s,\zeta_s)
\frac{\partial\mv a_{\rm T}^H(\chi_s,\zeta_s)}{\partial \zeta_s}$. 
The derivatives of the distance between the target and BS, and direction cosines with respect to the target horizontal coordinates are
$\mv g_{s,d}\triangleq[\frac{\partial d_s}{\partial x_s},\frac{\partial d_s}{\partial y_s}]^T=[\frac{x_s}{d_s},\frac{y_s}{d_s}]^T$,
$\mv g_{s,\chi}\triangleq[\frac{\partial \chi_s}{\partial x_s},\frac{\partial \chi_s}{\partial y_s}]^T=\big[\frac{y_s^2+\left(H_{\rm T}-H_{\rm BS}\right)^2}{d_s^3},-\frac{x_sy_s}{d_s^3}\big]^T$, and
$\mv g_{s,\zeta}\triangleq[\frac{\partial \zeta_s}{\partial x_s},\frac{\partial \zeta_s}{\partial y_s}]^T=\big[-\frac{\left(H_{\rm T}-H_{\rm BS}\right)x_s}{d_s^3},-\frac{\left(H_{\rm T}-H_{\rm BS}\right)y_s}{d_s^3}\big]^T$.
For $i\in\{1,2\}$, we define
\begin{align}
\mv D_{s,i}\triangleq[\mv g_{s,\chi}]_i\mv B_{s,\chi}+[\mv g_{s,\zeta}]_i\mv B_{s,\zeta}
-\frac{2[\mv g_{s,d}]_i}{d_s}\mv B_s.
\end{align}
It follows that $\frac{\partial(d_s^{-2}\mv B_s)}{\partial[\mv u_s]_i}=d_s^{-2}\mv D_{s,i}$. 
The first-order derivatives of $\boldsymbol{\mu}_{s,\ell}(\mv w_s)$ are then given by $\frac{\partial \boldsymbol{\mu}_{s,\ell}}{\partial [\mv u_s]_i} =\frac{\beta_0\alpha_s}{d_s^2}\mv D_{s,i}\mv s_{s,\ell}$, for $i\in\{1,2\}$, $\frac{\partial \boldsymbol{\mu}_{s,\ell}}{\partial \alpha_s^{\rm R}} =\frac{\beta_0}{d_s^2}\mv B_s\mv s_{s,\ell}$, and
$\frac{\partial \boldsymbol{\mu}_{s,\ell}}{\partial \alpha_s^{\rm I}}=j\frac{\beta_0}{d_s^2}\mv B_s\mv s_{s,\ell}$.

Since the noise covariance is independent of $\mv w_s$, the standard Fisher information matrix (FIM) formula for CSCG observations gives the instantaneous data FIM as
\begin{align}
[\mv J_{s}^{(\rm D)}(\mv w_s)]_{ij}
&\triangleq
-\mathbb E_{\mv y_s\mid\mv w_s,\mv y_{0:s-1}}\left[
\frac{\partial^2 \log p(\mv y_s\mid \mv w_s,\mv y_{0:s-1})}
{\partial w_{s,i}\partial w_{s,j}}
\right]\nonumber\\
&=\frac{2}{\sigma^2}
\Re\left\{
\sum_{\ell=1}^{L}
\left(
\frac{\partial \boldsymbol{\mu}_{s,\ell}}
{\partial w_{s,i}}
\right)^H
\frac{\partial \boldsymbol{\mu}_{s,\ell}}
{\partial w_{s,j}}
\right\}.
\label{eq:Jphi_entry}
\end{align}
According to the partition of $\mv w_s$ in \eqref{eq:def-wt}, the instantaneous data FIM is partitioned as
\begin{align}
\mv J_{s}^{(\rm D)}(\mv w_s)
=
\begin{bmatrix}
\mv J^{(\rm D)}_{s,uu}(\mv w_s)
&
\mv J^{(\rm D)}_{s,u\alpha}(\mv w_s)
\\
\big(\mv J^{(\rm D)}_{s,u\alpha}(\mv w_s)\big)^T
&
\mv J^{(\rm D)}_{s,\alpha\alpha}(\mv w_s)
\end{bmatrix}.
\label{eq:Jphi_block}
\end{align}
We next characterize the dependence of the instantaneous data FIM on $\mv W_s$. From the definition of the transmit sample covariance matrix, we have $\sum_{\ell=1}^{L}\mv s_{s,\ell}\mv s_{s,\ell}^{H}=L\mv W_s$. For ease of exposition, define $\mv C_s\triangleq\mv B_s^H\mv B_s$, $\mv R_{s,i}\triangleq\mv D_{s,i}^H\mv B_s$, $\mv S_{s,i}\triangleq\frac{1}{2}(\mv R_{s,i}+\mv R_{s,i}^H)$, $\mv A_{s,i}\triangleq\frac{1}{2j}(\mv R_{s,i}-\mv R_{s,i}^H)$, and $\mv H_{s,ij}\triangleq\frac{1}{2}(\mv D_{s,i}^H\mv D_{s,j}+\mv D_{s,j}^H\mv D_{s,i})$. Then, we can obtain, for $i,j\in\{1,2\}$, $\mv\Xi_{s,uu,ij}(\mv w_s)\triangleq\frac{2L|\alpha_s|^2}{\sigma^2}\,\frac{|\beta_0|^2}{d_s^4}\,\mv H_{s,ij}$, $\mv\Xi_{s,u\alpha^{\rm R},i}(\mv w_s)\triangleq\frac{2L|\beta_0|^2}{\sigma^2}\frac{1}{d_s^4}\big(\alpha_s^{\rm R}\mv S_{s,i}+\alpha_s^{\rm I}\mv A_{s,i}\big)$, and $\mv\Xi_{s,u\alpha^{\rm I},i}(\mv w_s)\triangleq\frac{2L|\beta_0|^2}{\sigma^2}\frac{1}{d_s^4}\big(\alpha_s^{\rm I}\mv S_{s,i}-\alpha_s^{\rm R}\mv A_{s,i}\big)$.

Based on the above definitions, the location information block is given by
\begin{align}
\hspace{-3.5mm}\mv J^{(\rm D)}_{s,uu}\!(\mv w_s)\!
\!=\!\!
\begin{bmatrix}
\!{\rm tr}(\mv\Xi_{s,uu,11}(\mv w_s)\mv W_s)
&
\!\!\!\!\!{\rm tr}(\mv\Xi_{s,uu,12}(\mv w_s)\mv W_s)\!
\\
\!{\rm tr}(\mv\Xi_{s,uu,21}(\mv w_s)\mv W_s)
&
\!\!\!\!\!{\rm tr}(\mv\Xi_{s,uu,22}(\mv w_s)\mv W_s)\!
\end{bmatrix}\!\!.\!\!\!\!\!
\label{eq:J_xx_pointwise}
\end{align}
The cross information block between the target location and the RCS coefficient is given by
\begin{align}
\hspace{-3.0mm}\mv J^{(\rm D)}_{s,u\alpha}\!(\mv w_s)\!
\!=\!\!\!
\begin{bmatrix}
\!{\rm tr}(\mv\Xi_{s,u\alpha^{\rm R},1}(\mv w_s)\mv W_s)
&
\!\!\!\!\!\!{\rm tr}(\mv\Xi_{s,u\alpha^{\rm I},1}(\mv w_s)\mv W_s)\!
\\
\!{\rm tr}(\mv\Xi_{s,u\alpha^{\rm R},2}(\mv w_s)\mv W_s)
&
\!\!\!\!\!\!{\rm tr}(\mv\Xi_{s,u\alpha^{\rm I},2}(\mv w_s)\mv W_s)\!
\end{bmatrix}\!\!.\!\!\!\!
\label{eq:J_xalpha_pointwise}
\end{align}
To express the RCS information block, define
$\mv\Xi_{s,\alpha\alpha}(\mv w_s)
\triangleq
\frac{2L}{\sigma^2}\,
\frac{|\beta_0|^2}{d_s^4}\,
\mv C_s.$
Since
$
\Re\!\big\{
\sum_{\ell=1}^{L}
\big(\frac{\partial \boldsymbol\mu_{s,\ell}}{\partial \alpha_s^{\rm R}}\big)^H
\big(\frac{\partial \boldsymbol\mu_{s,\ell}}{\partial \alpha_s^{\rm I}}\big)
\big\}
=0,$
the $(\boldsymbol\alpha_s,\boldsymbol\alpha_s)$ block is diagonal and is given by
\begin{align}
\mv J^{(\rm D)}_{s,\alpha\alpha}(\mv w_s)
=
{\rm tr}\big(\mv\Xi_{s,\alpha\alpha}(\mv w_s)\mv W_s\big)\,\mv I_2.
\label{eq:J_alphaalpha_pointwise}
\end{align}
The conditional data PFIM is obtained by averaging the instantaneous data FIM over the predictive PDF as
\begin{align}
\mv J^{(\rm D)}_{s\mid0:s-1}
=
\mathbb E_{\mv w_s\mid \mv y_{0:s-1}}\!\left[
\mv J_s^{(\rm D)}(\mv w_s)
\right], 
\label{eq:Jdata_exact_standard}
\end{align}
where the expectation is taken over the exact predictive PDF $p(\mv w_s\mid\mv y_{0:s-1})$. Conditioned on $\mv y_{0:s-1}$, $\mv W_s$ does not depend on $\mv w_s$, and thus $\mv J^{(\rm D)}_{s\mid 0:s-1}$ is affine in $\mv W_s$. 
Since the exact predictive PDF $p(\mv w_s\mid\mv y_{0:s-1})$ in (\ref{eq:predictive_density_exact}) generally has no closed-form expression, \eqref{eq:Jdata_exact_standard} is difficult to characterize. 
\vspace{-3mm}
\subsection{Gaussian Approximation and Prior Information Matrix}
\vspace{-2mm}
To obtain a tractable representation for the predictive PDF, we approximate the posterior PDF by a Gaussian information state. This representation summarizes the realized observation history through its mean and covariance and admits closed-form propagation under the linear Gaussian state evolution model. 
Accordingly, we approximate the posterior PDF as
\begin{align}
\hspace{-2mm}\widetilde p_{s-1\mid0:s-1}(\mv w_{s-1})
\triangleq
\mathcal N\!\left(
\mv w_{s-1};
\bar{\mv w}_{s-1\mid0: s-1},
\mv \Sigma_{s-1\mid0: s-1}
\right),\!\!\!
\label{eq:assumed_filtering_density}
\end{align}
where $\bar{\mv w}_{s-1\mid 0:s-1}$ and $\mv\Sigma_{s-1\mid 0:s-1}$ denote the mean and covariance of the posterior Gaussian information state, respectively.  Under the linear Gaussian transition model in \eqref{eq:motion_state}, propagating \eqref{eq:assumed_filtering_density} gives the assumed predictive PDF
\begin{align}
\widetilde p_{s\mid0:s-1}(\mv w_s)
&\triangleq
\int
p(\mv w_s\mid \mv w_{s-1})
\widetilde p_{s-1\mid0:s-1}(\mv w_{s-1})
d\mv w_{s-1}
\nonumber\\
&=
\mathcal N\!\left(
\mv w_s;
\bar{\mv w}_{s\mid0: s-1},
\mv \Sigma_{s\mid 0:s-1}
\right),
\label{eq:assumed_density_prediction}
\end{align}
where
\begin{align}
\bar{\mv w}_{s\mid 0:s-1}
&=
\mv F_{s-1}\bar{\mv w}_{s-1\mid0:s-1},
\label{eq:predicted_mean}\\
\mv \Sigma_{s\mid0: s-1}
&=
\mv F_{s-1}\mv \Sigma_{s-1\mid 0:s-1}\mv F_{s-1}^{T}
+
\mv Q^w_{s-1}.
\label{eq:predicted_cov}
\end{align}
Thus, the intractable predictive PDF $p(\mv w_s\mid\mv y_{0:s-1})$ is approximated by $\widetilde p_{s\mid0:s-1}(\mv w_s)$, in which the mean and covariance recursively incorporate the tracking observation history. Based on this assumed predictive PDF, the prior information matrix is approximated by
\begin{align}
\widetilde{\mv J}_{s\mid0:s-1}^{(\rm P)}
&\triangleq
-
\mathbb E_{\widetilde p_{s\mid 0:s-1}}\!\left[
\nabla_{\mv w_s}^2
\log
\widetilde p_{s\mid0:s-1}(\mv w_s)
\right].
\label{eq:Jprior_assumed_density_def}
\end{align}
Since $\widetilde p_{s\mid0:s-1}(\mv w_s)$ is Gaussian, the negative Hessian of its log density is constant, yielding
\begin{align}
\widetilde{\mv J}_{s\mid 0:s-1}^{(\rm P)}
=
\mv \Sigma_{s\mid 0:s-1}^{-1}.
\label{eq:Jprior_gaussian}
\end{align}
Conformably with the partition of $\mv w_s=[\mv u_s^T,\boldsymbol\alpha_s^T]^T$, the prior information matrix is partitioned as
\begin{align}
\widetilde{\mv J}_{s\mid 0:s-1}^{(\rm P)}
=
\begin{bmatrix}
\widetilde{\mv J}^{(\rm P)}_{s\mid0:s-1,uu}
&
\widetilde{\mv J}^{(\rm P)}_{s\mid 0:s-1,u\alpha}
\\
\widetilde{\mv J}^{(\rm P)}_{s\mid 0:s-1,\alpha u}
&
\widetilde{\mv J}^{(\rm P)}_{s\mid 0:s-1,\alpha\alpha}
\end{bmatrix}.
\label{eq:Jprior_block}
\end{align}
The off-diagonal blocks $\widetilde{\mv J}^{(\rm P)}_{s\mid0:s-1,u\alpha}$ and $\widetilde{\mv J}^{(\rm P)}_{s\mid0:s-1,\alpha u}$ may be nonzero because the posterior information update can introduce statistical dependence between the target location and RCS coefficient, which is retained during state prediction.

Since $\mv u_0$ and $\mv\alpha_0$ are independent at $s=0$, the initial assumed predictive PDF of $\mv w_0=[\mv u_0^T,\mv\alpha_0^T]^T$ is given by $\widetilde p_{0\mid-1}(\mv w_0)=\mathcal N(\mv w_0;\bar{\mv w}_{0\mid-1},\mv\Sigma_{0\mid-1})$, where $\bar{\mv w}_{0\mid-1}=[\bar{\mv u}_0^T,\mv 0_2^T]^T$ and $\mv\Sigma_{0\mid-1}=\mathrm{blkdiag}\big(\mv C_0,\frac{\sigma_\alpha^2}{2}\mv I_2\big)$. The corresponding prior information matrix is
\begin{align}
\widetilde{\mv J}_{0\mid-1}^{(\mathrm P)}=\mv\Sigma_{0\mid-1}^{-1}=\mathrm{blkdiag}\big(\mv C_0^{-1},\frac{2}{\sigma_\alpha^2}\mv I_2\big).
\label{eq:Jprior_initial_gaussian}
\end{align} 
At each time slot, the assumed predictive PDF $\widetilde p_{s\mid0:s-1}(\mv w_s)$ is used to construct the prior information for designing $\mv W_s$. 
After collecting $\mv y_s$, the posterior mean and covariance are updated using the maximum a posteriori (MAP) estimation and conditional PFIM, respectively, and then propagated to the next time slot. 
   
\subsection{Assumed Density Conditional PFIM and Posterior Update}

Under the assumed predictive PDF, the data information matrix is approximated as
\begin{align}
\widetilde{\mv J}^{(\rm D)}_{s\mid0:s-1}
\triangleq
\mathbb E_{\widetilde p_{s\mid 0:s-1}}
\!\left[
\mv J_s^{(\rm D)}(\mv w_s)
\right].
\label{eq:Jdata_assumed_standard}
\end{align}
For each coefficient matrix in \eqref{eq:J_xx_pointwise}-\eqref{eq:J_alphaalpha_pointwise}, we define
\begin{align}
\widehat{\mv\Xi}_{s,\kappa}\triangleq\mathbb E_{\widetilde p_{s\mid0:s-1}}\big[\mv\Xi_{s,\kappa}(\mv w_s)\big],
\end{align}
where $\kappa$ denotes its corresponding subscript. The resulting information blocks are given by 
\begin{align}
\widetilde{\mv J}^{(\rm D)}_{s\mid0:s-1,uu}
&=
\begin{bmatrix}
{\rm tr}\big(\hat{\mv\Xi}_{s,uu,11}\mv W_s\big)
&
{\rm tr}\big(\hat{\mv\Xi}_{s,uu,12}\mv W_s\big)
\\
{\rm tr}\big(\hat{\mv\Xi}_{s,uu,21}\mv W_s\big)
&
{\rm tr}\big(\hat{\mv\Xi}_{s,uu,22}\mv W_s\big)
\end{bmatrix},
\label{eq:Jxx_final}
\end{align}
\begin{align}
\widetilde{\mv J}^{(\rm D)}_{s\mid 0:s-1,u\alpha}
&=
\begin{bmatrix}
{\rm tr}\big(\hat{\mv\Xi}_{s,u\alpha^{\rm R},1}\mv W_s\big)
&
{\rm tr}\big(\hat{\mv\Xi}_{s,u\alpha^{\rm I},1}\mv W_s\big)
\\
{\rm tr}\big(\hat{\mv\Xi}_{s,u\alpha^{\rm R},2}\mv W_s\big)
&
{\rm tr}\big(\hat{\mv\Xi}_{s,u\alpha^{\rm I},2}\mv W_s\big)
\end{bmatrix},
\label{eq:Jxalpha_final}
\end{align}
\begin{align}
\widetilde{\mv J}^{(\rm D)}_{s\mid0:s-1,\alpha\alpha}
&=
{\rm tr}\big(
\hat{\mv\Xi}_{s,\alpha\alpha}\mv W_s
\big)\mv I_2.
\label{eq:Jalphaalpha_final}
\end{align}
Moreover, we can obtain $\widetilde{\mv J}^{(\rm D)}_{s\mid 0:s-1,\alpha u}=\big(\widetilde{\mv J}^{(\rm D)}_{s\mid0:s-1,u\alpha}\big)^T$.

Combining the prior and data information gives the assumed conditional PFIM 
\begin{align}
\widetilde{\mv J}_{s\mid 0:s-1}=\widetilde{\mv J}^{(\rm P)}_{s\mid0:s-1}+\widetilde{\mv J}^{(\rm D)}_{s\mid0:s-1}.
\end{align}
Treating the RCS coefficient as a nuisance parameter, the dependence of the equivalent information matrix on $\mv W_s$ is explicitly given in \eqref{eq:J_eq_x_full}. 
\begin{figure*}[!t]
\begin{align}
\widetilde{\mv J}_{s\mid0:s-1,u}
&=\widetilde{\mv J}^{(\rm P)}_{s\mid0:s-1,uu}+
\begin{bmatrix}
\mathrm{tr}\big(\hat{\mv \Xi}_{s,uu,11}\mv W_s\big) &
\mathrm{tr}\big(\hat{\mv \Xi}_{s,uu,12}\mv W_s\big)
\\
\mathrm{tr}\big(\hat{\mv \Xi}_{s,uu,21}\mv W_s\big) &
\mathrm{tr}\big(\hat{\mv \Xi}_{s,uu,22}\mv W_s\big)
\end{bmatrix}\!-\!
\Big(
\widetilde{\mv J}^{(\rm P)}_{s\mid 0:s-1,u\alpha}
\!+\!
\begin{bmatrix}
\mathrm{tr}\big(\hat{\mv \Xi}_{s,u\alpha^{\rm R},1}\mv W_s\big) &
\mathrm{tr}\big(\hat{\mv \Xi}_{s,u\alpha^{\rm I},1}\mv W_s\big)
\\
\mathrm{tr}\big(\hat{\mv \Xi}_{s,u\alpha^{\rm R},2}\mv W_s\big) &
\mathrm{tr}\big(\hat{\mv \Xi}_{s,u\alpha^{\rm I},2}\mv W_s\big)
\end{bmatrix}
\Big)
\times\nonumber\\
&\Big(
\widetilde{\mv J}^{(\rm P)}_{s\mid 0:s-1,\alpha\alpha}
+
\mathrm{tr}\big(\hat{\mv \Xi}_{s,\alpha\alpha}\mv W_s\big)\mv I_2
\Big)^{-1}
\Big(
\widetilde{\mv J}^{(\rm P)}_{s\mid 0:s-1,\alpha u}
+
\begin{bmatrix}
\mathrm{tr}\big(\hat{\mv \Xi}_{s,u\alpha^{\rm R},1}\mv W_s\big) &
\mathrm{tr}\big(\hat{\mv \Xi}_{s,u\alpha^{\rm I},1}\mv W_s\big)
\\
\mathrm{tr}\big(\hat{\mv \Xi}_{s,u\alpha^{\rm R},2}\mv W_s\big) &
\mathrm{tr}\big(\hat{\mv \Xi}_{s,u\alpha^{\rm I},2}\mv W_s\big)
\end{bmatrix}^{T}
\Big).
\label{eq:J_eq_x_full}
\end{align}\vspace{-6mm}
\end{figure*}
The corresponding conditional PCRB for estimating $\mv u_s$ is
\begin{align}
\!\!\!\!{\rm PCRB}_{s\mid0:s-1}\triangleq{\rm tr}\big(\left[\widetilde{\mv J}_{s\mid0:s-1}^{-1}
\right]_{1:2,1:2}\big)\!={\rm tr}\big(\widetilde{\mv J}_{s\mid0:s-1,u}^{-1}\big).\!\!\!
\label{eq:PCRB_trace_form}
\end{align}

 After applying $\mv W_s$ and collecting $\mv y_s$, we update the posterior Gaussian information state. For a given $\mv u_s$, the observation model is linear in $\boldsymbol\alpha_s$, and its conditional MAP estimation can be obtained in closed form as 
\begin{align}
&\!\!\!\widehat{\boldsymbol\alpha}_s(\mv u_s) \nonumber\\
&\!\!\!\!\!\!\triangleq\arg\max_{\boldsymbol\alpha_s}
\big\{\!\log p\big(\mv y_s\mid\mv u_s,\boldsymbol\alpha_s\big)\!+\!\log\widetilde p_{s\mid0:s-1}\!([\mv u_s^T,\mv\alpha_s^T]^T)\big\}
\end{align} 
\begin{align}&\!\!\!\!\!\!=\Big(\widetilde{\mv J}^{(\mathrm P)}_{s\mid 0:s-1,\alpha\alpha}
+\frac{2L|\beta_0|^2}{\sigma^2d_s^4}
\operatorname{tr}(\mv C_s\mv W_s)\mv I_2\Big)^{-1}\Big(
\frac{2}{\sigma^2}\nonumber\\
&\!\!\!\!\!\!\times
\big[\Re\big\{\dfrac{\beta_0^*}{d_s^2}\displaystyle\sum_{\ell=1}^{L}
\mv s_{s,\ell}^{H}\mv B_s^{H}\mv y_{s,\ell}\big\}, 
\Im\big\{\dfrac{\beta_0^*}{d_s^2}\displaystyle\sum_{\ell=1}^{L}
\mv s_{s,\ell}^{H}\mv B_s^{H}\mv y_{s,\ell}\big\}\big]^T\nonumber\\
&\!\!\!\!\!\!+\widetilde{\mv J}^{(\mathrm P)}_{s\mid 0:s-1,\alpha\alpha}
\overline{\boldsymbol\alpha}_{s\mid 0:s-1}
-\widetilde{\mv J}^{(\mathrm P)}_{s\mid 0:s-1,\alpha u}
\big(\mv u_s-\overline{\mv u}_{s\mid0:s-1}\big)
\Big), 
\label{eq:conditional_rcs_map}
\end{align} 
where $\bar{\mv u}_{s\mid 0: s-1}$ and $\bar{\boldsymbol\alpha}_{s\mid0: s-1}$ denote the target location and RCS coefficient of the predictive mean $\bar{\mv w}_{s\mid 0:s-1}=[\bar{\mv u}_{s\mid 0:s-1}^{T},\bar{\boldsymbol\alpha}_{s\mid0: s-1}^{T}]^{T}$ in \eqref{eq:predicted_mean}, respectively.
Substituting $\widehat{\boldsymbol\alpha}_s(\mv u_s)$ into the joint posterior objective yields
\begin{align}
\widehat{\mv u}_s^{\rm MAP}
\triangleq\arg\max_{\mv u_s}&
\big\{\log p\big(\mv y_s\mid\mv u_s,\widehat{\boldsymbol\alpha}_s(\mv u_s)\big)\nonumber\\
&\!\!\!+\log\widetilde p_{s\mid 0:s-1}
\left(
\begin{bmatrix}
\mv u_s^T,
\widehat{\mv\alpha}_s(\mv u_s)^T
\end{bmatrix}^T
\right)
\big\}.
\label{eq:concentrated_map}
\end{align}
 The posterior Gaussian information state is then specified by the mean
\begin{align}
\bar{\mv w}_{s\mid0: s}\triangleq\begin{bmatrix}(\widehat{\mv u}_s^{\rm MAP})^T,\big(\widehat{\boldsymbol\alpha}_s(\widehat{\mv u}_s^{\rm MAP})\big)^T\end{bmatrix}^T
\label{eq:posterior_mean_map}
\end{align}
and the covariance is approximated by the inverse of the conditional PFIM as
\begin{align}
\mv\Sigma_{s\mid0: s}\triangleq\left[\widetilde{\mv J}_{s\mid0: s-1}\right]^{-1}.
\label{eq:posterior_covariance_pfim}
\end{align}
The resulting information state is propagated through \eqref{eq:predicted_mean} and \eqref{eq:predicted_cov}, with the overall procedure summarized in Algorithm~\ref{alg:proposed_sequential_design}.

\section{Problem Formulation}\label{problem_formulation}

At each time slot $s=0,\ldots,S$, the transmit covariance $\mv W_s\succeq \mv 0$ is designed based on the assumed predictive PDF $\widetilde p_{s\mid0:s-1}(\mv w_s)$, which is initialized by the prior distributions at $s=0$ and recursively updated and predicted for $s\geq 1$. The design problem is formulated as
\begin{align}
(\mathrm{P1}\text{-}s)\quad
\min_{\mv W_s\succeq \mv 0}\quad
&
{\rm tr}\!\left(\widetilde{\mv J}_{s\mid 0:s-1,u}^{-1}\right)
\\
{\rm s.t.}\quad
&
{\rm tr}(\mv W_s)\leq P .
\end{align}
Problem $(\mathrm{P1}\text{-}s)$ is non-convex due to the matrix inverse involved in the objective function and the structure of $\widetilde{\mv J}_{s\mid 0:s-1,u}$. Nevertheless, by leveraging Schur complement, (P1-$s$) can be equivalently transformed into a convex semi-definite program (SDP), as detailed below.

\section{Optimal Solution to (P1-$s$)}\label{proposed_solution}
First, we introduce an auxiliary matrix $\mv \Psi_s\succeq \mv 0$ satisfying $\mv \Psi_s\preceq\widetilde{\mv J}_{s\mid 0:s-1,u}$. Since $\widetilde{\mv J}_{s\mid0:s-1,\alpha\alpha}\succ\mv 0$, the Schur complement represents this inequality equivalently as
\begin{align}
\begin{bmatrix}
\widetilde{\mv J}_{s\mid 0:s-1,uu}-\mv\Psi_s
&
\widetilde{\mv J}_{s\mid0:s-1,u\alpha}
\\
\widetilde{\mv J}_{s\mid 0:s-1,\alpha u}
&
\widetilde{\mv J}_{s\mid 0:s-1,\alpha\alpha}
\end{bmatrix}
\succeq \mv 0.
\label{eq:LMI_schur1}
\end{align}
where all blocks are affine in $(\mv W_s,\mv\Psi_s)$. 
Then, we transform the trace-inverse objective into an equivalent convex form. We introduce an auxiliary matrix $\mv Z_s\succeq\mv 0$ satisfying
$\mv Z_s\succeq\mv\Psi_s^{-1}$. 
Note that $\mv\Psi_s\succ\mv 0$ holds at the optimal solution. Based on the Schur complement, the above constraint is equivalent to 
\begin{align}
\begin{bmatrix}
\mv \Psi_s & \mv I_2\\
\mv I_2 & \mv Z_s
\end{bmatrix}
\succeq \mv 0.
\label{eq:LMI_schur2}
\end{align}
Problem $(\mathrm{P1}\text{-}s)$ can then be reformulated as
\begin{align}
(\mathrm{P1}\text{-}\mathrm{SDP}\text{-}s)\!\!\!
\min_{\mv W_s\succeq \mv 0,\mv \Psi_s\succeq \mv 0,\mv  Z_s\succeq \mv 0:{\rm tr}(\mv W_s)\le P,\eqref{eq:LMI_schur1},\eqref{eq:LMI_schur2}}
\quad
& {\rm tr}(\mv Z_s).
\label{eq:P3t_SDP}
\end{align}
Since $\widetilde{\mv J}^{(\rm D)}_{s\mid 0:s-1,uu}$, $\widetilde{\mv J}^{(\rm D)}_{s\mid0:s-1,u\alpha}$, and $\widetilde{\mv J}^{(\rm D)}_{s\mid 0:s-1,\alpha\alpha}$ are affine in $\mv W_s$, the constraint in \eqref{eq:LMI_schur1} is affine in $(\mv W_s,\mv\Psi_s)$, while that in \eqref{eq:LMI_schur2} is affine in $(\mv\Psi_s,\mv Z_s)$. Therefore, $(\mathrm{P1}\text{-}\mathrm{SDP}\text{-}s)$ is a convex SDP. 
It remains to establish the equivalence between $(\mathrm{P1}\text{-}\mathrm{SDP}\text{-}s)$ and $(\mathrm{P1}\text{-}s)$. First, for any feasible $\mv W_s$ to $(\mathrm{P1}\text{-}s)$, setting $\mv\Psi_s=\widetilde{\mv J}_{s\mid 0:s-1,u}(\mv W_s)$ and $\mv Z_s=\widetilde{\mv J}_{s\mid 0:s-1,u}^{-1}(\mv W_s)$ gives a feasible solution to $(\mathrm{P1}\text{-}\mathrm{SDP}\text{-}s)$ with the same objective value. Hence, the optimal value of $(\mathrm{P1}\text{-}\mathrm{SDP}\text{-}s)$ is no larger than that of $(\mathrm{P1}\text{-}s)$.  
Conversely, for any feasible $(\mv W_s,\mv\Psi_s,\mv Z_s)$ to $(\mathrm{P1}\text{-}\mathrm{SDP}\text{-}s)$, \eqref{eq:LMI_schur1} and \eqref{eq:LMI_schur2} imply $\mv Z_s\succeq\mv\Psi_s^{-1}\succeq\widetilde{\mv J}_{s\mid 0:s-1,u}^{-1}(\mv W_s)$, and thus ${\rm tr}(\mv Z_s)\geq{\rm tr}\big(\widetilde{\mv J}_{s\mid0:s-1,u}^{-1}(\mv W_s)\big)$. Since $\mv W_s$ is also feasible for $(\mathrm{P1}\text{-}s)$ the optimal value of (P1-SDP-$s$). Therefore, the two problems have the same optimal value, and the optimal $\mv W_s^\star$ obtained from (P1-SDP-$s$) is also optimal for $(\mathrm{P1}\text{-}s)$. 
\begin{algorithm}[!t]
\caption{Proposed Algorithm}
\label{alg:proposed_sequential_design}
\begin{algorithmic}[1]
\STATE Initialize $\widetilde p_{0\mid-1}(\mv w_0)$.
\FOR{$s=0,\ldots,S$}
\STATE Construct the conditional information matrices under $\widetilde p_{s\mid0:s-1}(\mv w_s)$.
\STATE Solve $(\mathrm{P1}\text{-}\mathrm{SDP}\text{-}s)$ to obtain $\mv W_s^\star$.
\STATE Apply $\mv W_s^\star$ and collect $\mv y_s$.
\STATE Compute $\widehat{\boldsymbol\alpha}_s(\mv u_s)$ and $\widehat{\mv u}_s^{\rm MAP}$ from \eqref{eq:conditional_rcs_map} and \eqref{eq:concentrated_map}.
\STATE Construct $\widetilde p_{s\mid0:s}(\mv w_s)$ using \eqref{eq:posterior_mean_map} and \eqref{eq:posterior_covariance_pfim}.
\IF{$s<S$}
\STATE Propagate $\widetilde p_{s\mid0:s}(\mv w_s)$ to obtain $\widetilde p_{s+1\mid0:s}(\mv w_{s+1})$.
\ENDIF
\ENDFOR
\end{algorithmic}
\end{algorithm}
\vspace{-2mm}
\section{Numerical Results}\label{numerical_results}
\begin{figure}[!t]
\centering
\includegraphics[width=0.92\linewidth]{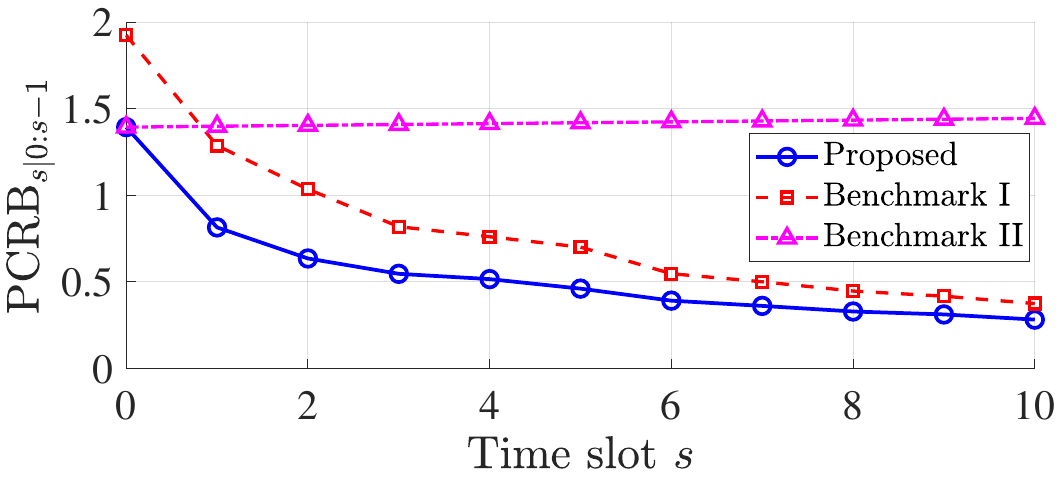}
\setlength{\abovecaptionskip}{-0.1cm}\vspace{2mm}
\caption{Conditional PCRB over sequential time slots.}
\label{fig_pcrb_vs_time} \vspace{-2mm}
\end{figure}

\begin{figure}[!t]
\centering
\includegraphics[width=0.92\linewidth]{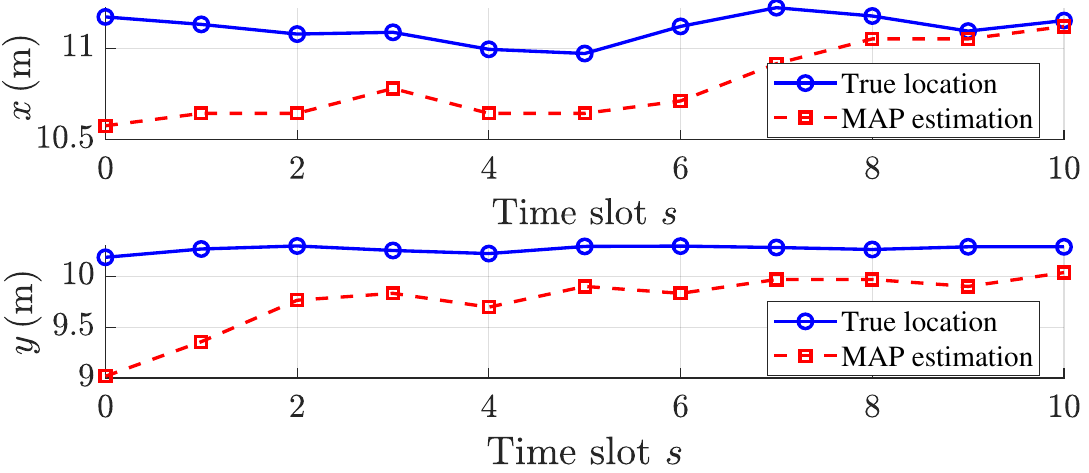}
\setlength{\abovecaptionskip}{-0.1cm}\vspace{2mm}
\caption{\mbox{Tracking results of the target horizontal location.}}
\label{Fig_xy_tracking} \vspace{-2mm}
\end{figure}

\begin{figure}[!t]
    \centering
    \vspace{-2mm}
    \includegraphics[width=0.94\columnwidth]{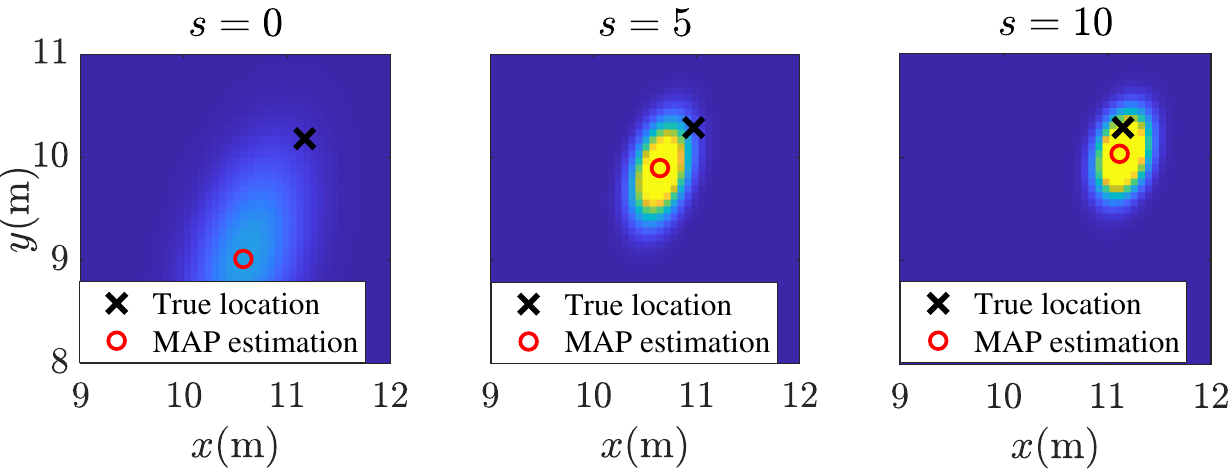}
    \vspace{-2mm}
    \caption{Posterior marginal distribution $\widetilde p(\mv u_s\mid \mv y_{0:s})$ and MAP estimate of the target location.}
    \label{Fig_predictive_density}
    \vspace{-2mm}
\end{figure}

\vspace{-2mm}
This section evaluates the proposed transmit covariance design under $S=10$ and
\(L=20\). We set
\(N_{\rm T}=N_{{\rm T},x}N_{{\rm T},z}=5\times4=20\) and
\(N_{\rm R}=N_{{\rm R},x}N_{{\rm R},z}=6\times4=24\). Unless otherwise specified, we set \(T_{\rm s}=100\) ms,
\(H_{\rm BS}=15\) m, \(H_{\rm T}=2\) m, \(\beta_0=-30\) dB,
\(\sigma^2=-90\) dBm, and \(P=20\) dBm. The initial target location follows
\(\mv u_0\sim\mathcal N(\bar{\mv u}_0,\mv C_0)\), with
\(\bar{\mv u}_0=[10,10]^T\) m and \(\mv C_0=\mv I_2~\mathrm{m}^2\).
For \(s\ge 1\), the target evolves according to the Gaussian random-walk
model with \(\mv Q_s=5\times10^{-3}\mv I_2~\mathrm{m}^2\). The RCS
coefficient is initialized as
\(\alpha_0\sim\mathcal{CN}(0,\sigma_\alpha^2)\), with
\(\sigma_\alpha^2=1\), and follows the Gauss-Markov model with
\(\rho=0.99\).
For comparison, we consider the following two benchmark schemes:
\begin{itemize}
    \item \textbf{Benchmark I: Isotropic covariance design.}
    The transmit covariance matrix is fixed as
    $\mv W_s^{\rm iso}=\frac{P}{N_{\rm T}}\mv I_{N_{\rm T}}$, for $s=0,\ldots,S$. This scheme does not optimize the transmit covariance matrix, and thus the probing strategy is not adapted to the realized observation history.

    \item \textbf{Benchmark II: History-independent covariance optimization.}
    This scheme solves the same transmit covariance optimization problem as the proposed design. The difference is that the proposed design constructs the assumed predictive PDF by recursively incorporating the observation history $\mv y_{0:s-1}$, whereas this benchmark forms the predictive PDF by only propagating the initial prior  distributions through the target location and RCS evolution models. Hence, the observation history is not exploited in the design of $\mv W_s$.
\end{itemize}
Fig.~\ref{fig_pcrb_vs_time} shows the conditional PCRB achieved by the proposed and benchmark schemes. At $s=0$, the proposed design and Benchmark II achieve the same PCRB since they use the same initial PDF. As $s$ increases, the PCRB of the proposed design decreases and remains lower than that of the benchmark schemes, which shows the benefits of transmit covariance optimization and observation history exploitation. 
Fig.~\ref{Fig_xy_tracking} compares the true and MAP estimated  locations of the target. As more tracking observations are collected, the MAP estimations gradually approach and follow the true target trajectory. This is consistent with the decreasing PCRB in Fig.~\ref{fig_pcrb_vs_time} and shows the progressive improvement in tracking. 
Fig.~\ref{Fig_predictive_density} presents the normalized posterior marginal distribution $\widetilde p(\mv u_s\mid\mv y_{0:s})$ at representative time slots. After collecting $\mv y_s$, the concentrated MAP mean update and the covariance update determine $\widetilde p_{s\mid 0: s}(\mv w_s)$. The target location marginal is obtained by integrating out the RCS coefficient as
$\widetilde p(\mv u_s\mid\mv y_{0:s})=\int\widetilde p_{s\mid 0:s}(\mv w_s)
d\boldsymbol\alpha_s$. 
As the tracking observations accumulate, it is observed that the posterior distribution becomes increasingly concentrated around the true target location, and the MAP estimate gradually follows the target trajectory.

\section{Conclusions}

This paper studied sequential transmit covariance design for target tracking in a MIMO radar system. We derived a Gaussian-assumed density conditional PCRB for the MSE in estimating the target's horizontal location as an explicit function. Based on this, we optimized the transmit covariance matrix to minimize the conditional PCRB at each time slot. 
The resulting PCRB minimization problem was equivalently transformed into a convex SDP via the Schur complement technique. Numerical results showed that the proposed design effectively exploits historical observations, reduces posterior uncertainty, and improves tracking accuracy over time. 

\vspace{-2mm}
\bibliographystyle{IEEEtran}
\bibliography{Tracking_GC} 
\end{document}